\documentclass[twocolumn,showpacs,pra]{revtex4}
\usepackage{graphicx}
\usepackage{bm}
\usepackage{amssymb}
\begin{document}

\title{Formation of molecules from a Cs Bose-Einstein condensate}

\author{V. A. Yurovsky}

\author{A. Ben-Reuven}

\affiliation{School of Chemistry, Tel Aviv University, 69978 Tel Aviv,
Israel}

\date{\today}
\begin{abstract}

Conversion of an expanding Bose-Einstein condensate of Cs atoms to a
molecular one with an efficiency of more than 30\% was observed
 recently
in experiments by M. Mark {\it et al.,} Europhys. Lett. {\bf 69}, 706
 (2005).
The theory  presented here describes the experimental results. Values
 of
resonance strength of 8 mG  and rate coefficients for atom-molecule
deactivation of $1\times 10^{-11}$  cm$^{3}/$s and molecule-molecule
 one of $1.5\times 10^{-9}$
cm$^{3}/$s are estimated by a fit of the theoretical results to the
experimental data. Near the resonance, where the highest conversion
efficiency was observed, the results demonstrate strong sensitivity to
the magnetic field ripple and inhomogeneity. A conversion efficiency
 of
about 60\% is predicted by non-mean-field calculations for the
 densities
and sweep rates lower than the ones used in the experiments.

\end{abstract}
\maketitle

\section*{Introduction}

A molecular Bose-Einstein condensate (BEC) has been recently
formed in experiments on atomic BEC
\cite{HKMWCNG03,MKHCNG04,DVMR03,DVMR04,XMACMK03,MAXCK04} and on
quantum-degenerate Fermi gases \cite{RTBJ03}. The molecules have been
formed by sweeping the Zeeman shift through a Feshbach resonance (see
Ref.\ \cite{TTHK99}) in a backward direction, so that the molecular
state crossed the atomic ones downwards. This led to the transfer of
population from the lowest atomic state in the case of a BEC, or from
an energy band in the case of a Fermi gas, to the molecular state, as
had been proposed in Ref.\ \cite{MTJ00}. Assuming all the atomic
population is initially in the BEC state, the backward sweep would
have been ideally suitable for forming molecules, were it not for two
loss mechanisms. The resonant molecule is generally populated in an
excited rovibrational and electronic state, and therefore can be
deactivated by exoergic inelastic collisions with atoms and other
molecules (see Refs.\ \cite{TTHK99,YBJW99,AV99,YBJW00}). (A
stabilization of molecules by means of coherent control \cite{SB03}
has been considered in Ref.\ \cite{LKMV04}.) In addition, during the
backward sweep, some higher-lying non-condensate atomic states can be
populated temporarily due to molecular dissociation. These two
effects restrict the efficiency of conversion from the atomic BEC to
the molecular one.

Formation of molecular BEC from degenerate Fermi gases, realized in
experiments \cite{RTBJ03}, is more efficient due to Pauli blocking of
inelastic collisions \cite{P03}. Some peculiarities of this process
 have
been discussed in Ref.\ \cite{PVB04}. In the case of Bose atoms the
deactivation losses can be minimized by a reduction of the condensate
density (see Ref.\  \cite{YB03}). Such a reduction can be realized in
experiments by using an expanding BEC, released from a trap by
 turning it
off (see Refs.\ \cite{HKMWCNG03,MKHCNG04,DVMR03,DVMR04}). A conversion
efficiency of more than 30\%, comparable to the one obtained for Fermi
gases, has been achieved by this way in Cs experiments
 \cite{MKHCNG04}.

The present work provides a theoretical description of molecular
formation in an expanding BEC with applications to the Cs experiments
\cite{MKHCNG04}, including an explanation of the high conversion
efficiency. Some of the results obtained below can be satisfactory
derived by a mean-field theory. However, in general, a description of
certain processes, such as the spontaneous dissociation of the
molecular BEC into non-condensate entangled atom pairs discussed in
Sec.\ \ref{SecMeth} below requires the use of a non-mean-field
theory. Several theoretical methods are available for this purpose.
One such method is based on a numerical solution of stochastic
differential equations in the positive-$P$ representation, as used in
the present context in Ref.\ \cite{PM01}. Another method is the
Hartree-Fock-Bogoliubov formalism (see Refs.\ \cite{HPW01,KH02}),
which deals with coupled equations for the atomic and the molecular
mean fields, as well as the normal and the anomalous densities
describing the second-order correlations of the non-condensate atomic
fields. These correlations are also taken into account in the
microscopic quantum dynamics approach used in Ref.\ \cite{KB02}. Some
of these methods, however, have no room for incorporating the
deactivating collisions. The parametric approximation used in Refs.\
\cite{YB03,VYA01,YB04,Y05} incorporates both the non-mean-field
effects and the damping due to deactivating collisions.

The present work is organized in the following way. Section
\ref{SecMeth} describes the various processes in a hybrid
 atom-molecule
condensate and theoretical methods needed for their analysis. The
 necessary
parameters of the Cs BEC are estimated in Sec.\ \ref{SecParEst} by a
 fit of
calculation results to the experimental data. The formation of
 molecules in
the switching scheme, one of the two sweeping methods used in the
experiment \cite{MKHCNG04} and discussed in Sec.\ \ref{SecParEst}
 below, is
analyzed in Sec.\ \ref{SecSwitch}. Optimal conditions for the
 molecular
formation are determined in Sec.\ \ref{SecOpt}.

\section{Theoretical methods\label{SecMeth}}

The effect of Feshbach resonance appears in a BEC of Cs atoms when the
collision energy of a pair of atoms in an open channel is close to the
energy of a bound state Cs$_{2}\left( m\right) $ in a closed channel
(see Ref.\
\cite{TTHK99}). The temporary formation and dissociation of the
 resonant
(Feshbach) molecular state Cs$_{2}\left( m\right) $ can be described
 as a reversible reaction
\begin{equation}
\text{Cs + Cs  }\rightleftarrows\text{ Cs}_{2}\left( m\right)  .
 \label{RCol}
\end{equation}
The Cs$_{2}\left( m\right) $ is an excited rovibrational state with
 orbital angular
momentum $l=4$, belonging to an excited state of the fine and
 hyperfine
structures. This state can be deactivated by an exoergic collision
with a third atom of the condensate \cite{TTHK99,YBJW99,AV99,YBJW00},
\begin{equation}
\mathrm{Cs}_{2}(m)+\text{Cs}\rightarrow \text{Cs}_{2}(d)+\text{Cs  ,
 }\label{AMCol}
\end{equation}
bringing the molecule down to a lower state Cs$_{2}\left( d\right) $,
 and
releasing kinetic energy to the relative motion of the
reaction products. Although the collision occurs with a
vanishingly small kinetic energy, rates of such inelastic
processes remain finite at near-zero energies
\cite{MAXCK04,Forrey99}. A variant of this process, involving
deactivation by a collision with another molecule (rather than
an atom), of the type
\begin{equation}
\mathrm{Cs}_{2}(m)+\mathrm{Cs}_{2}(m)\rightarrow \text{Cs}_{2}\left(
 d\right) +\text{Cs}_{2}\left( u\right)  , \label{MMCol}
\end{equation}
would require a significant molecular density to be
effective. The two molecular states Cs$_{2}\left( d\right) $ and
 Cs$_{2}\left( u\right) $ can be
distinct.

A complete analysis of the processes in a hybrid atom-molecule BEC
 must
take into account both the relaxation processes due to deactivating
collisions (\ref{AMCol}) and (\ref{MMCol}), and the quantum
 fluctuations due
to dissociation of the resonant molecules to non-condensate atoms
(\ref{RCol}). A starting point of such an analysis can be the quantum
equation of motion for the atomic field annihilation operator
 $\hat{\Psi }_{a}\left( {\bf p},t\right) $ in the
momentum representation. An adiabatic elimination of the ``dump''
 states
Cs$_{2}\left( d\right) $ and Cs$_{2}\left( u\right) $ (see Refs.\
 \cite{YB03,Y05}) reduces this equation to the
Heisenberg-Langevin stochastic equation
\begin{eqnarray}
i\hbar \dot{\hat{\Psi }}_{a}\left( {\bf p},t\right)  =\left\lbrack
 {p{ } ^{2}\over 2m} + \epsilon _{a}\left( t\right) -i {k{ }
 _{a}\over 2}|\varphi _{m}\left( t\right) |^{2}\right\rbrack
 \hat{\Psi }_{a}\left( {\bf p},t\right)  \nonumber
\\
+2g^{*}\varphi _{m}\left( t\right) \hat{\Psi }^{\dag }_{a}\left(
 -{\bf p},t\right)  +i \hat{F}\left( {\bf p},t\right)  . \label{Psia}
\end{eqnarray}
Here $m$ is the atomic mass, $\epsilon _{a}\left( t\right) =-{1\over
 2}\mu \left( B\left( t\right) -B_{0}\right) $ is the time-dependent
Zeeman shift of the atom in an external magnetic field $B\left(
 t\right) $ relative to
half the energy of the molecular state (which is fixed as the
 zero-energy
point), $\mu $ is the difference between the magnetic momenta of an
 atomic pair
and a molecule, and $B_{0}$  is the resonance value of $B$. The
 atom-molecule
hyperfine coupling $g$ is related to the phenomenological resonance
 strength
$\Delta $ through $|g|^{2}=2\pi \hbar ^{2}|a_{a}|\mu \Delta /m$ (see
 Ref.\ \cite{YBJW00}), where $a_{a}$  is the
background elastic scattering length for atom-atom collisions. The
deactivation (\ref{AMCol}) by atom-molecule collisions is represented
 in
Eq.\ (\ref{Psia}) by the imaginary term, proportional to the
 deactivation
rate $k_{a}$, as well as by the quantum noise source $\hat{F}\left(
 {\bf p},t\right) $, related by a
fluctuation-dissipation theorem. The quantum noise is required in
 order to
maintain the correct commutation relations of the atomic field
 operators.

In the parametric approximation \cite{YB03,Y05}, the
quantum fluctuations of the molecular field are neglected, and
the molecules are described by a mean field $\varphi _{m}\left(
 t\right) $. The atomic
field operator is expressed in this method as
\begin{equation}
\hat{\Psi }_{a}\left( {\bf p},t\right) =C\left( t\right) \left\lbrack
 \hat{A}\left( {\bf p},t\right) \psi _{c}\left( p,t\right)
+\hat{A}^{\dag }\left( -{\bf p},t\right) \psi _{s}\left( p,t\right)
 \right\rbrack , \label{PsiaA}
\end{equation}
where the damping factor $C\left( t\right) $ takes into account the
 imaginary
term in Eq.\ (\ref{Psia}),
\begin{equation}
C\left( t\right) =\exp\left( -\int\limits^{t}_{t{ } _{0}}d t_{1}{k{ }
 _{a}\over 2}|\varphi _{m}\left( t_{1}\right) |^{2}\right)  ,
\end{equation}
and the $c$-number functions $\psi _{c,s}\left( p,t\right) $ satisfy
 the ordinary
differential equations (with $p$ as a parameter)
\begin{equation}
i\hbar \dot{\psi }_{c,s}\left( p,t\right) =\left\lbrack {p{ }
 ^{2}\over 2m} + \epsilon _{a}\left( t\right) \right\rbrack  \psi
 _{c,s}\left( p,t\right) +2g^{*}\varphi _{m}\left( t\right) \psi ^{
*}_{s,c}\left( p,t\right)  . \label{Psics}
\end{equation}
The initial conditions $\psi _{c}\left( p,t_{0}\right) =1$, $\psi
 _{s}\left( p,t_{0}\right) =0$ are introduced at
$t=t_{0}$, assuming the atomic field is then a coherent state of zero
kinetic energy. The operators $\hat{A}\left( {\bf p},t\right) $ can
 be expressed in terms of the
functions $\psi _{c,s}\left( p,t\right) $ and the quantum noise
 $\hat{F}\left( {\bf p},t\right) $. As a result of
ensuing analysis (see Refs.\ \cite{YB03,Y05}), the atomic density
comprises the sum
\begin{equation}
n_{a}\left( t\right) =n_{0}\left( t\right) +n_{s}\left( t\right)
 \label{adens}
\end{equation}
of the densities of condensate atoms
\begin{equation}
n_{0}\left( t\right) =|\varphi _{0}\left( t\right) |^{2}, \label{n0}
\end{equation}
and of non-condensate (entangled) atoms
\begin{equation}
n_{s}\left( t\right) =\left( 2\pi \hbar \right) ^{-3}\int d^{3}p
 n_{s}\left( p,t\right)  .
\end{equation}
Here
\begin{equation}
\varphi _{0}\left( t\right) =C\left( t\right) \left\lbrack \psi
 _{c}\left( 0,t\right) \varphi _{0}\left( t_{0}\right) +\psi
 _{s}\left( 0,t\right) \varphi ^{*}_{0}\left( t_{0}\right)
 \right\rbrack  \label{phi0p}
\end{equation}
is the atomic condensate mean field. The momentum spectrum of the
non-condensate atoms
\begin{eqnarray}
n_{s}\left( p,t\right) =|\psi _{s}\left( p,t\right) |^{2}\left\lbrack
 1+\eta _{s}\left( p,t\right) \right\rbrack +|\psi _{c}\left(
 p,t\right) |^{2}\eta _{s}\left( p,t\right)  \nonumber
\\
-2\text{Re}\left\lbrack \psi ^{*}_{s}\left( p,t\right) \psi _{c}\left
( p,t\right) \eta _{c}\left( p,t\right) \right\rbrack  , \label{ns}
\end{eqnarray}
as well as their anomalous density [encountered in Eq.\
(\ref{Phimr}) below]
\begin{eqnarray}
m_{s}\left( p,t\right) =\psi _{s}\left( p,t\right) \psi _{c}\left(
 p,t\right) \left\lbrack 1+2\eta _{s}\left( p,t\right) \right\rbrack
 \nonumber
\\
-\psi ^{2}_{c}\left( p,t\right) \eta _{c}\left( p,t\right) -\psi
 ^{2}_{s}\left( p,t\right) \eta ^{*}_{c}\left( p,t\right)  ,
\end{eqnarray}
are expressed in terms of the auxiliary functions
\begin{eqnarray}
\eta _{s}\left( p,t\right) =k_{a}C^{2}\left( t\right)
 \int\limits^{t}_{t{ } _{0}}{d t^\prime \over C^{2}\left( t^\prime
 \right) } |\varphi _{m}\left( t^\prime \right) \psi _{s}\left(
 p,t^\prime \right) |^{2} \nonumber
\\
{}\label{etacs}
\\
\eta _{c}\left( p,t\right) =k_{a}C^{2}\left( t\right)
 \int\limits^{t}_{t{ } _{0}}{d t^\prime \over C^{2}\left( t^\prime
 \right) }|\varphi _{m}\left( t^\prime \right) |^{2}\psi _{s}\left(
 p,t^\prime \right) \psi ^{*}_{c}\left( p,t^\prime \right)  ,
 \nonumber
\end{eqnarray}
which describe the contribution of quantum noise.

The equation of motion for the molecular mean field has the form
(see Refs.\ \cite{YB03,Y05})
\begin{eqnarray}
i\hbar \dot{\varphi }_{m}\left( t\right) =g \varphi ^{2}_{0}\left(
 t\right) -i\left( {k{ } _{a}\over 2}n_{a}\left( t\right) +k_{m}
|\varphi _{m}\left( t\right) |^{2}\right) \varphi _{m}\left( t\right)
  \nonumber
\\
+{1\over 2\pi ^{2}\hbar { } ^{3}}\int\limits^{\infty }_{0}d p
 \left\lbrack p^{2}g m_{s}\left( p,t\right) +2\hbar m|g|^{2}\varphi
 _{m}\left( t\right) \right\rbrack , \label{Phimr}
\end{eqnarray}
where $k_{m}$  is the rate coefficient of the molecule-molecule
deactivating collisions (\ref{MMCol}). The second term under the
integral over $p$ appears as a result of a renormalization procedure
(see Refs.\ \cite{HPW01,Y05}), necessary in order to regularize the
integral. A numerical solution of Eqs.\ (\ref{Psics}) on a grid of
values of $p$, combined with Eq.\ (\ref{Phimr}), is consistently
sufficient for elucidating the dynamics of the system.

The parametric approximation considered above is particularly
suitable for the analysis of homogeneous systems. It can be applied
also to inhomogeneous systems using a local density approximation, but
its application to a strongly inhomogeneous expanding BEC meets
serious difficulties. Fortunately, under proper conditions this case
can be treated sufficiently well by a mean field approach (see Ref.\
\cite{YB04}), neglecting the atomic field quantum fluctuations. The
applicability of this simpler approach can be verified by a comparison
of results of the parametric and mean-field calculations for the
corresponding homogeneous system.

The expansion of a pure atomic BEC has been considered in Ref.\
\cite{expan} by the introduction of scaled normal coordinates
\begin{equation}
\rho _{j}=r_{j}/b_{j}\left( t\right) , 1\le j\le 3,
\end{equation}
where the scales $b_{j}$  obey the equations
\begin{equation}
\ddot{b}_{j}\left( t\right) =\omega ^{2}_{j}/\left\lbrack b_{1}\left(
 t\right) b_{2}\left( t\right) b_{3}\left( t\right) b_{j}\left(
 t\right) \right\rbrack  , \label{exp_b}
\end{equation}
in which the $\omega _{j}$  are the angular frequencies of the
 harmonic trap
containing the condensate before expansion. The initial conditions
$b_{j}\left( t_{\exp}\right) =1$, $\dot{b}_{j}\left( t_{\exp}\right)
 =0$ are stated at the start of the expansion $t_{\exp}$.
Solutions of Eq.\ (\ref{exp_b}) (see Ref.\ \cite{expan}) demonstrate
that the expansion is ballistic after an acceleration period of
$\sim \min\left( \omega ^{-1}_{j}\right) $. As shown in Ref.\
 \cite{YB04}, the molecules  inherit the
velocity of the atoms they are formed from. The atomic and molecular
mean fields can be represented in terms of rescaled fields
$\Phi _{0}\left( \bm{\rho },t\right) $ and $\Phi _{m}\left( \bm{\rho
 },t\right) $, respectively, as
\begin{eqnarray}
\varphi _{0}\left( {\bf r},t\right) =A\left( t\right) \Phi _{0}\left(
 \bm{\rho },t\right) e^{i}{ } ^{S} \nonumber
\\
\label{phi}
\\
\varphi _{m}\left( {\bf r},t\right) =A\left( t\right) \Phi _{m}\left(
 \bm{\rho },t\right) e^{2i}{ } ^{S} , \nonumber
\end{eqnarray}
where the scaling factor $A\left( t\right) =\left( b_{1}\left(
 t\right) b_{2}\left( t\right) b_{3}\left( t\right) \right) ^{-1/2}$
 describes
the density reduction, and the phase factor with
\begin{equation}
S\left( t\right)  ={m\over \hbar }
 \sum\limits^{3}_{j=1}r^{2}_{j}{\dot{b}_{j}\left( t\right) \over
 2b_{j}\left( t\right) }-{\epsilon { } _{0}\over \hbar }
 \int\limits^{t}_{t{ } _{\exp}}dt^\prime A^{2}\left( t^\prime \right)
\end{equation}
contains most of the contribution of the kinetic energy. Here
$\epsilon _{0}=4\pi \hbar ^{2}a_{a}n_{0}/m$ is a chemical potential
 of the atomic BEC and $n_{0}$  is its
peak density while the trap is on. As a result ( see Ref.\
\cite{YB04}), the rescaled mean fields obey the set of ordinary
differential equations
\begin{eqnarray}
i\hbar \dot{\Phi }_{0}\left( \bm{\rho },t\right) =\left\lbrack
 \epsilon _{a}\left( t\right) -{i\over 2}A^{2}\left( t\right) k_{a}
|\Phi _{m}\left( \bm{\rho },t\right) |^{2}\right\rbrack \Phi
 _{0}\left( \bm{\rho },t\right)  \nonumber
\\
+2A\left( t\right) g^{*}\Phi ^{*}_{0}\left( \bm{\rho },t\right) \Phi
 _{m}\left( \bm{\rho },t\right)  \nonumber
\\
\label{GPexp}
\\
i\hbar \dot{\Phi }_{m}\left( \bm{\rho },t\right) =-iA^{2}\left(
 t\right) \biggl\lbrack {1\over 2}k_{a}|\Phi _{0}\left( \bm{\rho
 },t\right) |^{2} \nonumber
\\
+k_{m}|\Phi _{m}\left( \bm{\rho },t\right) |^{2}\biggr\rbrack \Phi
 _{m}\left( \bm{\rho },t\right) +A\left( t\right) g\Phi ^{2}_{0}\left
( \bm{\rho },t\right)  . \nonumber
\end{eqnarray}
The coordinate dependence arises from the use of inhomogeneous
Thomas-Fermi initial conditions for $\Phi _{0}\left( \bm{\rho
 },t_{\exp}\right) $, while
$\Phi _{m}\left( \bm{\rho },t_{\exp}\right) =0$.

\section{Parameter estimates\label{SecParEst}}

Molecules have been formed in the experiments \cite{MKHCNG04} by
using a very weak Feshbach resonance in Cs situated near 20 G. The
resonance strength and the rate coefficients of atom-molecule and
molecule-molecule deactivation are unknown and are estimated here by a
fit of the calculation results to the experimental data.

The magnetic field has been varied in these experiments in two
manners. In the {\it ramping} scheme the magnetic field has been swept
through resonance with a fixed ramp speed. In the {\it switching}
 scheme the
magnetic field has been tuned to a value $B_{\text{test}}$  in the
 vicinity of the
resonance and then held for a fixed time $t_{\text{hold}}$, starting
 from the
value $B_{\text{start}}=B_{0}\pm 0.5$ G for $B_{\text{test}}\gtrless
 B_{0}$, respectively, so that the
resonance should not be crossed. Due to finite response time the
magnetic field variation  is represented by an exponential function,
\begin{equation}
B\left( t\right) =B_{\text{test}}+\left(
 B_{\text{start}}-B_{\text{test}}\right) \exp\left\lbrack \left(
 t-t_{\exp}\right) /1.54\text{ ms}\right\rbrack  .
\end{equation}
The switching scheme has been applied both to the trapped and the
expanding BEC. In the last case, the magnetic field variation has been
started from $B_{\text{start}}$  simultaneously with the expansion at
 $t=t_{\exp}$.

\begin{figure}
\includegraphics[width=3.375in]{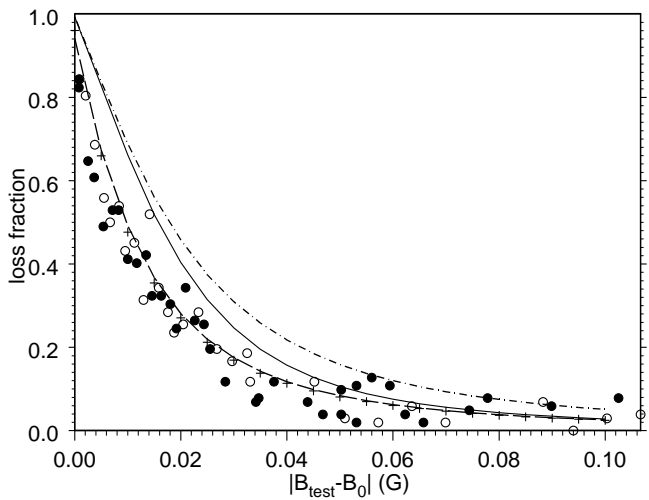}

\caption{Switching-scheme atom loss fraction calculated for the
trapped BEC with $\Delta =8$ mG, $k_{m}=1.5\times 10^{-9}$  cm$^{3}
/$s,  $k_{a}=1\times 10^{-11}$  cm$^{3}/$s (solid
line) and  $k_{a}=2\times 10^{-11}$  cm$^{3}/$s (dot-dashed line).
 The results for
$k_{m}=1\times 10^{-10}$  cm$^{3}/$s are calculated with $\Delta =8$
 mG, $k_{a}=1\times 10^{-11}$  cm$^{3}/$s
(dashed line) and $\Delta =2$ mG, $k_{a}=4\times 10^{-11}$  cm$^{3}
/$s (pluses). The open and
solid circles represent the experimental data of Ref.\
\protect\cite{MKHCNG04} measured below and above the resonance,
respectively.} \label{fig_btesthd}

\end{figure}

Consider first the {\it switching} scheme for a {\it trapped} BEC,
resulting in a condensate loss with a negligible molecular
formation. This experiment is similar to the slow-sweep Na
experiments \cite{MIT_Na_loss} and the $^{87}$Rb experiments
 \cite{M02},
in which the field was stopped short of resonance, too. In those
cases the condensate loss is determined by simple analytical
expressions involving the product of the resonance strength $\Delta $
 and
the atom-molecule deactivation rate coefficient $k_{a}$  (see Refs.\
\cite{YBJW99,YBJW00,YB03b}). In the present case, these analytical
expressions are inapplicable owing to the non-linear magnetic field
variation. However, numerical calculations still demonstrate a
dependence on the product $k_{a}\Delta $ provided moderate
 molecule-molecule
deactivation rates are assumed. (This is demonstrated by the dashed
line and pluses in Fig.\ \ref{fig_btesthd}). The comparison with the
experimental data leads to an estimated value of $k_{a}\Delta \approx
 8\times 10^{-11}$  mG
cm$^{3}/$s. Much higher values of $k_{m}$  do not lead to such a good
 fit.
Although the condensate loss becomes dependent on a variation of
 $k_{a}$
and $\Delta $, keeping the product $k_{a}\Delta $ fixed, this
 variation does not
improve the fit.

\begin{figure}
\includegraphics[width=3.375in]{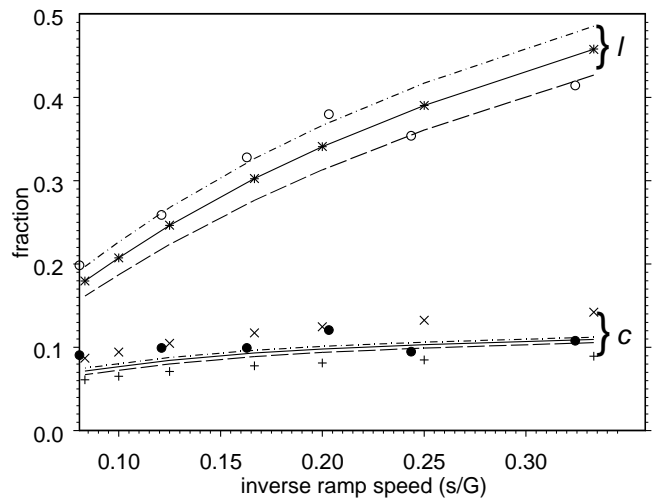}

\caption{Ramping-scheme atom loss fraction (l) and conversion
efficiency (c) calculated for the expanding BEC with $k_{m}=1.5\times
 10^{-9}$
cm$^{3}/$s,  $k_{a}\Delta =8\times 10^{-11}$  mG cm$^{3}/$s, and
 $\Delta =8$ mG (solid lines), $\Delta =7$ mG
(dashed lines), or $\Delta =9$ mG (dot-dashed lines). The results for
 $\Delta =8$ mG
and $k_{a}=1\times 10^{-11}$  cm$^{3}/$s are represented by pluses
($k_{m}=2\times 10^{-9}$  cm$^{3}/$s) and
crosses ($k_{m}=1\times 10^{-9}$  cm$^{3}/$s). The open and solid
 circles represent the
experimental data of Ref.\ \protect\cite{MKHCNG04} for loss and
conversion, respectively.} \label{fig_ramp}

\end{figure}

Consider now the {\it ramping} scheme. It has been applied in
 experiments
\cite{MKHCNG04} to the expanding BEC, measuring both the numbers of
remained atoms and formed molecules. In the fast-decay approximation
(see
Refs.\ \cite{YBJW99,YBJW00}), the atomic condensate loss due to the
resonance crossing is determined by the resonance strength $\Delta $
 only and is
independent of the deactivation rates. Although the analytical
 expressions
of Refs.\ \cite{YBJW99,YBJW00} are inapplicable to the case of an
 expanding
BEC, the results of numerical calculations demonstrate a low
 sensitivity of
the BEC loss to the deactivation rates. A comparison with the
 experimental
data leads to the estimated value of $\Delta \approx 8$ mG  (see the
 upper graphs in Fig.\
\ref{fig_ramp}). Together with the above estimate for $k_{a}\Delta $
 this leads to
$k_{a}\approx 1\times 10^{-11}$  cm$^{3}/$s. The experimental data
 points presented in Fig.\
\ref{fig_ramp} were obtained with a magnetic field ramp that has been
started at $t_{\exp}$,  using a magnetic field value such that the
 resonance is
crossed in 10 ms and the populations are measured 20 ms after
 $t_{\exp}$
\cite{ChinPC}.

The remaining unknown parameter, the molecule-molecule deactivation
rate coefficient $k_{m}$, can be estimated by a comparison of the
 calculation
results with the experimental data for the number of atoms converted
 to
molecules (see lower graphs in Fig.\ \ref{fig_ramp}). This leads to
 the
estimated value of $k_{m}\approx 1.5\times 10^{-9}$  cm$^{3}/$s. The
 graphs in Fig.\ \ref{fig_ramp}
are practically insensitive to $k_{a}$  because of its much lower
 value. A look
back at Fig.\ \ref{fig_btesthd} for the switching scheme shows that,
 at
most, $k_{a}$  is bounded by $k_{a}<2\times 10^{-11}$  cm$^{3}/$s,
 when the far-off-resonance
results are regarded. In this range, the results are not much
 sensitive to
the value of $k_{m}$. However, nearer resonance, the higher value of
 $k_{m}$  would
lead to an overestimate of the loss, independently of $k_{a}$  and
 $\Delta $.

Let us compare these values with what is known regarding other atoms.
The estimated value of the rate coefficient of atom-molecule
 deactivation
$k_{a}\approx 10^{-11}$  cm$^{3}/$s is several times lower than the
 corresponding values of
$5.5\times 10^{-11}$  cm$^{3}/$s for Na measured in Ref.\
 \cite{MAXCK04} and $7\times 10^{-11}$  cm$^{3}/$s
for $^{87}$Rb estimated in Ref.\ \cite{YB03b}. A molecule-molecule
 deactivation
rate coefficient $k_{m}\approx 1.5\times 10^{-9}$  cm$^{3}/$s exceeds
 the corresponding value for Na
$2.5\times 10^{-11}$  cm$^{3}/$s  measured in Ref.\ \cite{MAXCK04} by
 two orders of
magnitude. These differences may be related to the large orbital
 angular
momentum ($l=4$) of the resonant molecular state in the present Cs
 case,
while for Na and $^{87}$Rb $l=0$. But there may be another reason.

All the estimates made here concerning the data for molecular
conversion are based on the suggestion that the resonant beam,
 blasting out
the atoms in the experiments \cite{MKHCNG04}, does not affect the
molecules. (This procedure was used to separate the atoms from
 molecules.)
If some part of the molecular population is removed by the blasting
 pulse,
the experimental results can be explained using lower values of
 $k_{m}$.

\begin{figure}
\includegraphics[width=3.375in,clip]{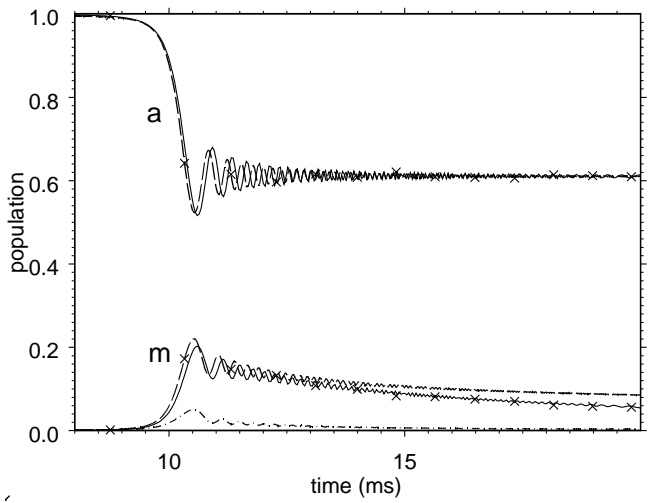}

\caption{Time dependence of the atomic  (a) and molecular (m)
condensate populations (scaled to the initial atomic one) calculated
in a ramping scheme for a homogeneous BEC with the initial atomic
density $6.4\times 10^{11}$  cm$^{-3}$, ramp speed 3 G/s (in the
 backward sweep), $\Delta =8$
mG, $k_{a}=1\times 10^{-11}$  cm$^{3}/$s, and $k_{m}=1\times 10^{-9}$
  cm$^{3}/$s with the parametric (solid
lines) and mean-field (crosses) approximations. The dashed lines
represent the mean field results for the expanding BEC with the
initial atomic density $7.6\times 10^{13}$  cm$^{-3}$. The
 non-condensate atom
population calculated with the parametric approximation is plotted by
the dot-dashed line.} \label{fig_tdramp3}

\end{figure}

All the calculations above were performed with the mean field
approximation. Figure \ref{fig_tdramp3} compares results of the
 parametric
and mean-field calculations for the initial atomic density of
 $6.4\times 10^{11}$
cm$^{-3}$, corresponding to the mean density at the resonance
 crossing for the
slowest ramp speed of 3 G/s used in the experiments \cite{MKHCNG04}.
 This
figure demonstrates that the temporary non-condensate atom population
persists only about 1 ms and after this short time the results of
parametric and mean-field calculations for the homogeneous case
 coincide
within a good accuracy. The expansion reduces the deactivation losses
compared to the ones for the homogeneous case.

\section{Molecular formation in the switching scheme\label{SecSwitch}}

The highest conversion efficiency, reaching beyond 30\%, has been
observed in the experiments \cite{MKHCNG04} for the switching scheme
 in an
{\it expanding} BEC. The characteristic time of atom-molecule
 relaxation $t_{am}$  is
determined by the coupling terms in Eq.\ (\ref{GPexp}) as
\begin{equation}
t^{-1}_{am}\sim |Ag\Phi _{0}|=\left( {2\pi |a_{a}|\mu \Delta \over
 m}n_{0}\right) ^{1/2} ,
\end{equation}
where $n_{0}\left( t\right) =|\varphi _{0}\left( t\right) |^{2}$  is
 non-rescaled atomic condensate density. Even
for $n_{0}=10^{12}$  cm$^{-3}$, corresponding to the resonance
 approach time in the
experiments \cite{MKHCNG04}, $t_{am}\sim 1$ ms is less than the
 characteristic time
of the magnetic field variation (1.54 ms). Therefore the evolution of
 the
atom-molecule condensate is adiabatic. Neglecting the deactivation,
 it can
be described by a quasi-stationary solution of the coupled
 Gross-Pitaevskii
equations (see Ref.\ \cite{TTHK99})
\begin{eqnarray}
n_{0}={n\over 36}\left( \epsilon \sqrt{\epsilon ^{2}+24}-\epsilon
 ^{2}+24\right)  \nonumber
\\
\label{StatSol}
\\
n_{m}={n\over 144}\left( 2\epsilon ^{2}-2\epsilon \sqrt{\epsilon ^{2}
+24}+24\right)  , \nonumber
\end{eqnarray}
corresponding to the pure atomic BEC ($n_{m}=0$) above the resonance
 at
$\epsilon \rightarrow \infty $. Here $n_{m}=|\varphi _{m}|^{2}$  is
 the non-rescaled molecular density, $n=n_{0}+2n_{m}$  is the
total density of atoms, and
\begin{equation}
\epsilon ={2\epsilon { } _{a}\over |g|\sqrt{n}} ={B_{0}-B{ }
 _{\text{test}}\over \delta B} ,\qquad \delta B=\left( {2\pi \hbar
 ^{2}|a_{a}|\Delta \over m\mu }n_{0}\right) ^{1/2}
\end{equation}
is a dimensionless detuning (generally time-dependent). The later fast
sweep of the magnetic field going under the resonance, used in the
switching scheme, conserves the atomic and molecular densities
(\ref{StatSol}) acquired at $B=B_{\text{test}}$.

The conversion efficiency reaches its maximum
\begin{equation}
2{n{ } _{m}\over n}={1\over 3} \label{maxconveff}
\end{equation}
when the magnetic field levels off at $\epsilon =0$. In the case of
 large
detunings $|B_{\text{test}}-B_{0}|\gg \delta B$ the molecular density
 decreases as
$n_{m}\sim n\delta B^{2}/\left( B_{\text{test}}-B_{0}\right) ^{2}$.
 Therefore the substantial molecular population and
condensate losses induced by the deactivating collisions can take
 place
only while $|B_{\text{test}}-B_{0}|<\delta B\approx 0.2$ mG for
 $n=10^{12}$  cm$^{-3}$.

\begin{figure}
\includegraphics[width=3.375in]{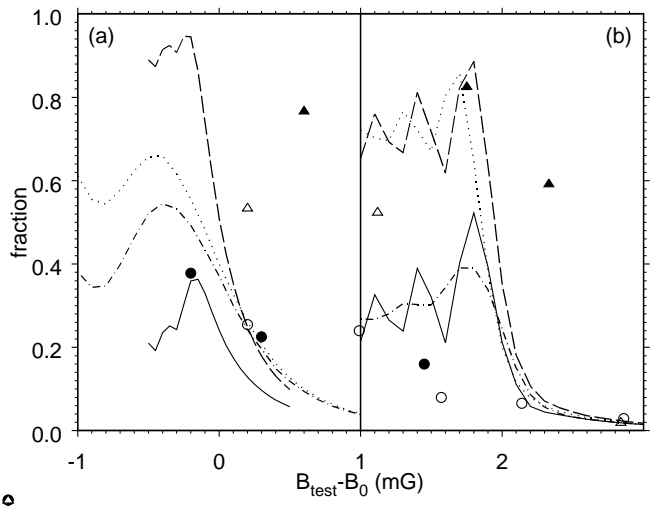}

\caption{(a) Atom loss fraction (dashed line) and conversion
efficiency (solid line) calculated as functions of the resonance
detuning for the switching scheme in the expanding BEC with
$k_{m}=1.5\times 10^{-9}$  cm$^{3}/$s,  $k_{a}=1\times 10^{-11}$
 cm$^{3}/$s, and $\Delta =8$ mG. The open and solid
circles represent the experimental data of Ref.\
\protect\cite{MKHCNG04} for loss and conversion, respectively. The
results of calculations for a rippled magnetic field with the phase
$\chi =1.25\pi $ are plotted by dot-dashed (conversion) and dashed
(loss)
lines. (b) The loss (dashed line) and conversion (solid line) for
$\chi =1.85\pi $. The results for inhomogeneous magnetic field with
 gradient 1
G/cm are plotted by dot-dashed (conversion) and dashed (loss) lines.
In both parts the triangles and circles represent the experimental
data of Ref.\ \protect\cite{MKHCNG04} for loss and conversion,
respectively.} \label{fig_btest}

\end{figure}

\begin{figure}
\includegraphics[width=3.375in]{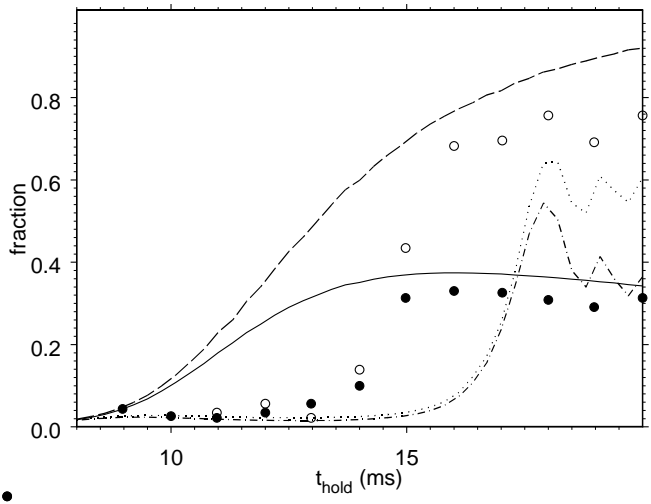}

\caption{Time dependence of the atom loss fraction (dashed line) and
conversion efficiency (solid line) calculated for the switching
 scheme in
the expanding BEC with $\Delta =8$ mG, $k_{m}=1.5\times 10^{-9}$
 cm$^{3}/$s,  $k_{a}=1\times 10^{-11}$  cm$^{3}/$s, and
$B_{\text{test}}=-0.15$ mG. The open and solid circles represent the
 experimental data
of Ref.\ \protect\cite{MKHCNG04} for loss and conversion,
 respectively. The
results of calculations for a rippled magnetic field with the phase
 $\chi =1.25\pi $
and $B_{\text{test}}=-0.4$ mG are plotted by dot-dashed (conversion)
 and dashed (loss)
lines.  \label{fig_tdswitch}}

\end{figure}

This conclusion is confirmed by the results of the numerical
calculations (see Fig.\ \ref{fig_btest}). These results, however,
 predict a
maximal conversion efficiency of 36\% for
 $B_{\text{test}}=B_{0}-0.15$ mG, when the
resonance is crossed. The smooth time dependence of the atomic and
 molecular
populations (see Fig.\ \ref{fig_tdswitch}) is in agreement with the
adiabatic evolution mentioned above. On decreasing further
 $B_{\text{test}}$  the
conversion efficiency decreases, demonstrating Rabbi oscillations due
 to
non-adiabatic effects. The results in Fig.\ \ref{fig_btest}
 correspond to a
resonance approach from above ($B_{\text{start}}=B_{0}+0.5$ G). In
 the mean field
approximation used here, neglecting molecular dissociation with
 formation of
non-condensate atoms, the same results are reached by approaching the
resonance from below.

Although the theory describes the peak conversion efficiency and
condensate losses observed in the experiments \cite{MKHCNG04}, the
 actual
experimental width of the resonance in conversion and in loss, of 2
 mG, is
about an order of magnitude more than the theoretical one. This
disagreement can be related to magnetic field variation mentioned in
 Ref.\
\cite{MKHCNG04}.

An ambient magnetic field ripple with a frequency of 50 Hz  and an
amplitude of 4 mG  leads to a quite perceptible effect. Although the
experiment has been synchronized with the ripple, the results of
calculations demonstrate a strong dependence on the ripple phase. In
 these
calculations the magnetic field time dependence has the form
\begin{eqnarray}
B\left( t\right) =B_{\text{test}}+\left(
 B_{\text{start}}-B_{\text{test}}\right) \exp\left\lbrack \left(
 t-t_{\exp}\right) /1.54\text{ ms}\right\rbrack  \nonumber
\\
+2\text{ mG }\sin\left\lbrack 2\pi \times 50\text{ Hz}(t-t_{\exp})
+\chi \right\rbrack  . \label{Bosc}
\end{eqnarray}
The variation of the phase $\chi $ leads to a shift of the peak and
changes the shape of the magnetic field dependence of the loss and
molecule formation (see Fig.\ \ref{fig_btest}).

\begin{figure}
\includegraphics[width=3.375in]{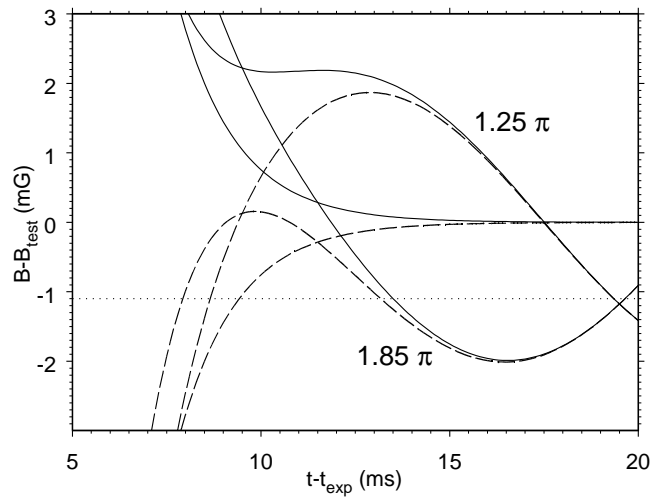}

\caption{Time dependence of the rippled magnetic field with the phases
$\chi =1.25\pi $, $\chi =1.85\pi $ and without ripples for
 $B_{\text{start}}=0.5$ G (solid lines) and
$B_{\text{start}}=-0.5$ G (dashed lines). An arbitrary value of
 $B_{0}$  is shown here for
reference by the dotted line.} \label{fig_bosc}

\end{figure}

\begin{figure}
\includegraphics[width=3.375in]{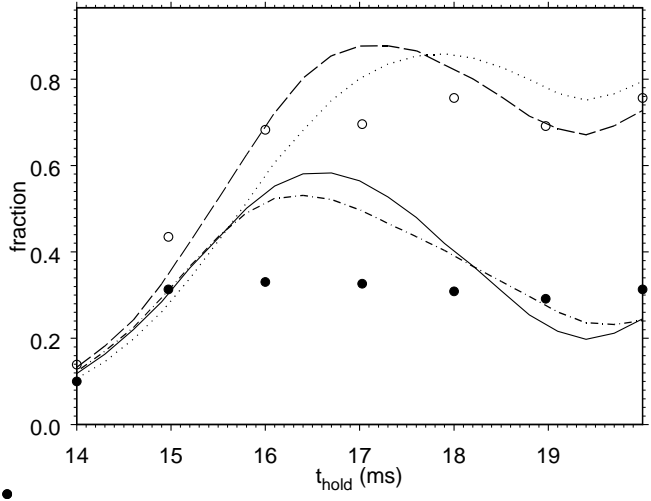}

\caption{Time dependence of the atom loss fraction (dashed
line) and conversion efficiency (solid line) calculated for a
rippled magnetic field with the phase $\chi =1.85\pi $, and with
 $B_{\text{test}}=1.7$
mG, $k_{m}=1.5\times 10^{-9}$  cm$^{3}/$s,  $k_{a}=1\times 10^{-11}$
 cm$^{3}/$s, and $\Delta =8$ mG. The results
for an inhomogeneous magnetic field with gradient 1 G/cm are plotted
by dot-dashed (conversion) and dashed (loss) lines.}
\label{fig_tdosc}

\end{figure}

The ripple may lead to the formation of molecules due to the
resonance crossing in a backward direction (see Fig.\ \ref{fig_bosc}).
The crossing can be non-adiabatic, leading to an oscillating time
dependence of the atomic and molecular populations (see Figs.\
\ref{fig_tdswitch} and \ref{fig_tdosc}). The oscillations are
 sensitive
to the ripple phase $\chi $ as well. For some values of $\chi $, e.
 g. $\chi =1.85$, the
resonance can be crossed a second time, but in the forward direction,
even though the first crossing occurred by approaching from above
($B_{\text{start}}>B_{0}$), given the arbitrary value of $B_{0}$
 shown in Fig.\
\ref{fig_bosc}. In this case, a rather long hold-on-time between the
crossing and measurement leads to sharp Rabbi oscillations. These
oscillations, however, can be averaged by a magnetic field
 inhomogeneity
with a rather small gradient of 1 mG/cm (see Figs.\ \ref{fig_btest}
 and
\ref{fig_tdosc}). Effects of both oscillations and inhomogeneity can
broaden the resonance to about 1 mG, which is still less than the
experimentally observed value of 2 mG. The additional broadening can
 be
related to the uncontrolled magnetic field variations of about 1 mG
mentioned in Ref.\ \cite{MKHCNG04}, the behavior of which is unclear.

The magnetic-field time dependence (\ref{Bosc}) can lead to different
results in approaching the resonance from below, with
 $B_{\text{start}}<B_{0}$, due to an
additional resonance crossing occurring earlier. This forward
 crossing can
lead to additional condensate losses due to molecular dissociation
 into
pairs of non-condensate atoms. This dissociation can, however, be
reversible, as the non-condensate atoms can associate to molecules
 during
the following backward crossing. A correct analysis of these effects
requires the application of non-mean-field calculations to the
 expanding
BEC, in order to obtain reliable conversion estimates.

\begin{figure}
\includegraphics[width=3.375in]{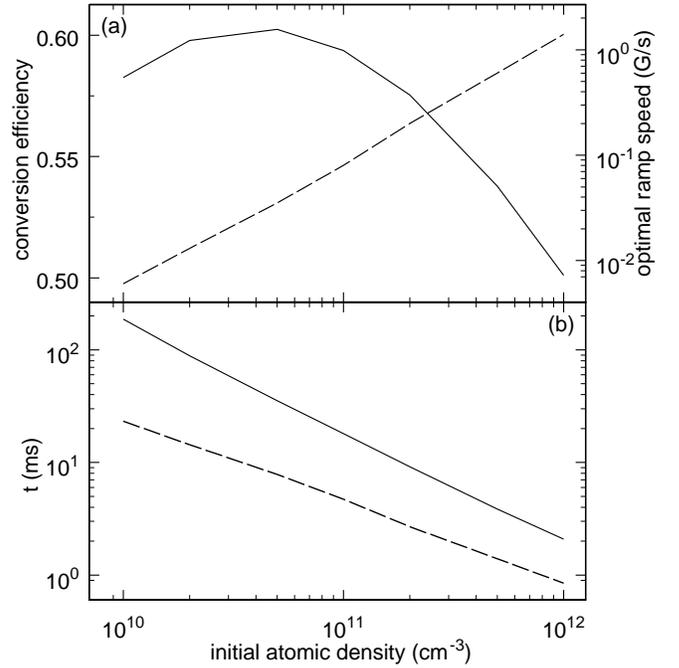}

\caption{(a) Conversion efficiency (solid line) at the optimal
ramp speed (dashed line). (b) The lifetime of the molecular condensate
(solid line) and the time after the resonance crossing when the peak
molecular density is reached (dashed line), using the appropriate
optimal ramp speed.\label{fig_opt}}

\end{figure}
\begin{figure}
\includegraphics[width=3.375in]{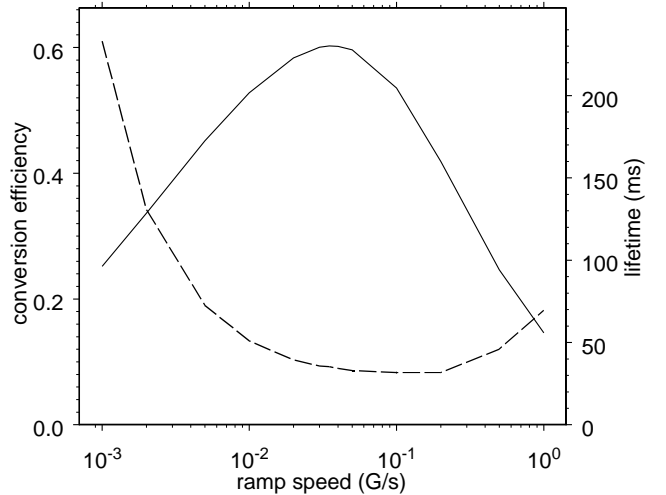}

\caption{Conversion efficiency (solid line) and the lifetime of
the molecular condensate (dashed line) as a function of the ramp speed
for the initial atomic density $5\times 10^{10}$
 cm$^{-3}$.\label{fig_rspeed}}

\end{figure}
\begin{figure}
\includegraphics[width=3.375in,clip]{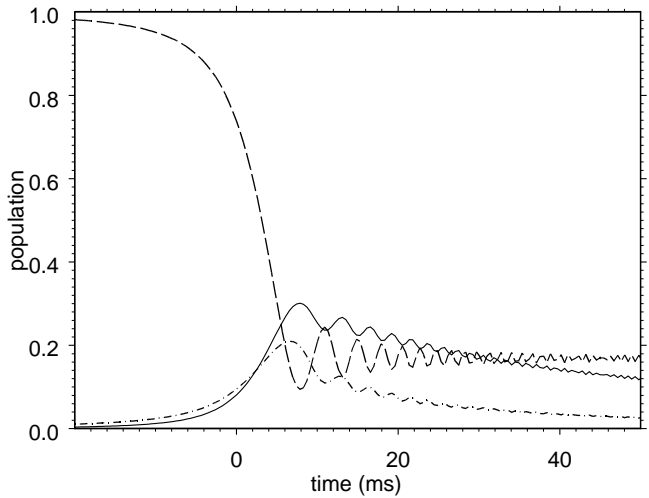}

\caption{Time dependence of the atomic  (dashed line) and
molecular (solid line) condensate populations (scaled to the initial
atomic one) calculated for a homogeneous BEC with the initial atomic
density $5\times 10^{10}$  cm$^{-3}$  and ramp speed 35 mG/s (in the
 backward sweep).
The non-condensate atom population is plotted by the dot-dashed line.
The resonance is crossed at $t=0$.} \label{fig_tdopt}

\end{figure}

\section{Optimal conditions\label{SecOpt}}

Having a model consistent with the experimental data, we can proceed
to determine optimal conditions for the molecular formation. In the
switching scheme, when the magnetic field adiabatically approaches the
resonance and further suddenly crosses it, the conversion efficiency
 is
restricted by the value of ${1\over 3}$ (see Eq.\ (\ref{maxconveff}).
 An adiabatic
crossing, as in the ramping scheme, allows for higher results since
 the
adiabatic state (\ref{StatSol}) corresponds to a total conversion
($n_{m}=n/2$)
in the limit of $\epsilon \rightarrow \infty $ below the resonance.
 As in the cases of Na \cite{YB03}
and $^{87}$Rb \cite{YB04,Y05}, the optimal ramp speed is determined
 by a balance
of the atomic association, decreasing at faster sweeps, and
 deactivation
losses, increasing at slower sweeps. Thus, a sudden sweep would lead
 to no
association, while an infinitely slow adiabatic sweep would lead to
 total
association, accompanied by a total deactivation loss during the
 infinite
time. The optimal density is determined by a concurrence of
 deactivation
losses, increasing at high densities, and dissociation into
 non-condensate
atoms, increasing at low densities. An analysis taking account of the
latter loss process requires a non-mean-field theory, such as the
parametric approximation. It was mentioned in Sec.\ \ref{SecMeth}
 above
that this approximation has its limitations in dealing with expanding
gases. However, the effect of increasing density due to expansion is
 less
important at the rather low optimal density, stated below, compared
 to the
conditions pertaining to Fig.\ \ref{fig_tdramp3}. For these reasons
 the
optimal conditions are determined by using the parametric
 approximation for
a homogeneous non-expanding BEC.

The results presented in Fig.\ \ref{fig_opt} show an optimal initial
atomic density of $5\times 10^{10}$  cm$^{-3}$.  This value
 corresponds to the mean density
of a Thomas-Fermi distribution with a peak density of $1.25\times
 10^{11}$  cm$^{-3}$.
Under the conditions of the experiments \cite{MKHCNG04}, this density
 can
be reached after 35 ms of expansion. The optimal ramp speed of 35 mG
/s is
much slower than the one used in the experiments. The high conversion
efficiency does not change much when the ramp speed is varied by
 about a
order of magnitude (see Fig.\ \ref{fig_rspeed}). Figure
 \ref{fig_tdopt}
shows that the molecular density reaches its maximum in about 8 ms
 after
the resonance crossing, or about 0.25 mG below the resonance, and the
 ramp
should be started about 2 mG above it. This figure demonstrates also a
substantial population of non-condensate atoms, justifying the
 necessity to
use non-mean-field calculations.

\section*{Conclusions}

The loss of Cs atomic BEC and formation of a molecular BEC observed in
the experiments \cite{MKHCNG04} can be described by a mean-field
 theory of
an expanding atom-molecule BEC. A fit of the calculation results to
 the
experimental data leads to estimated values of the resonance strength
 and
rate coefficients for atom-molecule and molecule-molecule deactivating
collisions. At small detunings the results are sensitive to magnetic
 field
ripple and inhomogeneity. A determination of optimal conditions for
 the
molecular formation requires a non-mean-field parametric
 approximation,
taking into account the dissociation of molecules into non-condensate
atomic pairs. A conversion efficiency of 60\% is predicted for lower
densities and slower sweeps than the ones used in the experiments.

\begin{acknowledgments}

The authors are very grateful to R. Grimm, H.-C. Nagerl, and C. Chin
for helpful discussions and clarification of experimental details.

\end{acknowledgments}

\end{document}